\documentclass[twocolumn,showpacs,amsmath,amssymb,prl]{revtex4}
\usepackage{graphicx}
\usepackage{dcolumn}
\usepackage{bm}
\begin{document}
\title{Harmonic lattice behavior of two-dimensional colloidal crystals}
\author{P. Keim, G. Maret, U. Herz and H.H. von Gr\"unberg}
\affiliation{Universit\"at Konstanz, Fachbereich Physik, P.O.B. 5560,
  78457 Konstanz, Germany}
\date{\today}
\begin{abstract}
  Using positional data from video-microscopy and applying the
  equipartition theorem for harmonic Hamiltonians, we determine the
  wave-vector-dependent normal mode spring constants of a
  two-dimensional colloidal model crystal and compare the measured
  band-structure to predictions of the harmonic lattice theory. We
  find good agreement for both the transversal and the longitudinal
  mode. For $q\to 0$, the measured spring constants are consistent
  with the elastic moduli of the crystal.
\end{abstract}
\pacs{61.66.-f,63.20.Dj,82.70.Dd}
\maketitle
Colloidal crystals have a long tradition as condensed matter analogs
of ordinary solids. They are studied, for example, to understand
phenomena such as freezing and melting \cite{naegele}.  Unlike in
ordinary solids having properties that are often difficult to connect
to the underlying atomic interactions, the interparticle potentials in
such colloidal crystals are in most cases precisely known and, more
importantly, externally controllable. Moreover, the relevant time- and
length scales in colloidal systems are comparatively easy to access
experimentally. Both aspects suggest studies directly probing
the connection between microscopic interaction-potentials and
macroscopic crystal properties.

The property we here consider is the crystal's elastic response to
thermal excitations, specifically, the phonon dispersion relations. In
this regard, colloidal crystals are rather special in that their
phonons are almost always overdamped: the ratio between the
wave-vector dependent frequency $\omega(\vec{q}) =
\sqrt{\lambda(\vec{q})/m}$, characteristic of the harmonic forces with
spring constants $\lambda(\vec{q})$, and the friction factors
$\Lambda(\vec{q})$ (also $q$-dependent \cite{hurd82}) for the modes of
lattice motion through the host liquid, is typically of the order
$10^{-3}$ to $10^{-4}$ in colloidal systems. Therefore, the time
autocorrelation function of a phonon normal mode coordinate decays
exponentially with a rate given by $\lambda(\vec{q})/\Lambda(\vec{q})$
\cite{hurd82,schram99}.  This decay rate can be, and has repeatedly
been, measured by means of dynamical light scattering
\cite{hurd82,piazza91,derksen92,hoppen98,cheng00} or inelastic light
scattering \cite{penciu02}. Phonon dispersion relations have been
determined in charge-stabilized \cite{hurd82,derksen92,hoppen98} and
purely hard sphere colloidal crystals \cite{cheng00,penciu02}, in the
context of dusty plasma physics \cite{schram93}, but also in more
exotic systems such as crystals made of mm steel spheres \cite{hay03}
or optically anisotropic spheres \cite{piazza91}.

Microscopic information about the spring constants and thus the
particle interaction potentials can only be derived from these decay
rates, i.e. from $\lambda(\vec{q})/\Lambda(\vec{q})$, if one resorts
to a model describing the complicated frictional forces, especially
those of hydrodynamic origin. A direct access to $\lambda(\vec{q})$,
i.e. one free from any assumptions of a model, is not possible in this
approach.

In this Letter we report on a video-microscopy study of
two-dimensional (2D) colloidal crystals and show how to obtain direct
access to the normal mode band-structure $\lambda(\vec{q})$ of the
crystal, circumventing, in particular, the difficulties arising from
the hydrodynamic interactions. The central idea is to avoid a
dynamical measurement and to analyze instead spatial correlations
between the particles which are then related to the
$\lambda(\vec{q})$, the eigenvalues of the dynamical matrix
characterizing the elastic properties of the harmonic crystal.  This
becomes possible through use of digital video-microscopy
\cite{murray96} providing us with the trajectories of all particles.
The colloidal system we examine is well-studied and the interparticle
potential precisely known \cite{zahn99,zahn00,zahn031}.  This will be
of advantage when establishing a quantitative link between the
measured $\lambda(\vec{q})$ and the theoretical band-structure based
on the pair-potential.

\begin{figure}
\includegraphics[width=0.45 \textwidth]{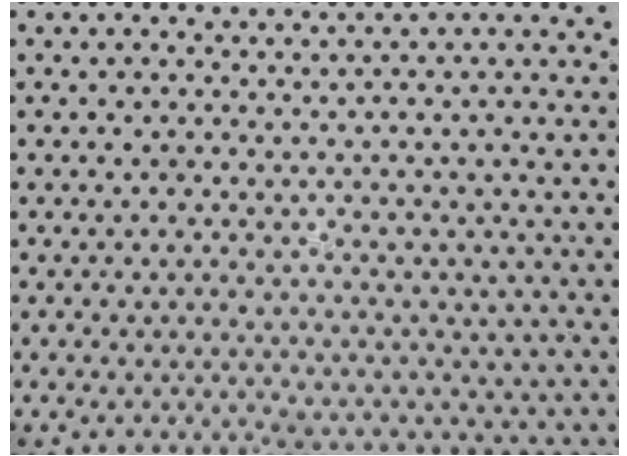}
\caption{\label{fig1}
  Micrograph ($420 \times 310\: \mu m$) of a typical colloidal crystal
  investigated in the present work; the two-dimensional system
  consists of paramagnetic colloids confined at the air-water
  interface of a hanging water drop.}
\end{figure}

The experimental setup is an improved version of the one described in
\cite{zahn99}: Spherical colloids (diameter $d=4.5 \:\mu m$) are
confined by gravity to a water/air interface formed by a water drop
suspended by surface tension in a top sealed cylindrical hole of a
glass plate. The flatness of the interface can be controlled within
+/- half a micron. The field of view has a size of $835 \times 620 \:
\mu m$ containing typically up to $3\cdot10^{3}$ particles, whereas
the whole sample contains about $10^{5}$ particles.  The number of
particles in the field of view is controlled by the curvature of the
droplet via an active regulation with an accuracy of 1\% and the
biggest observed particle-density gradient is 0.7\%. The variation of
the inclination of the sample is in the range of $\alpha \approx
1\:\mu$Rad so that the collective motion of the particles relative to
the field of view is below $2\:\mu m/h$ providing best equilibrium
conditions for long time stability.  The particles are
super-paramagnetic due to Fe$_{2}$O$_{3}$ doping. A magnetic field
$\vec{B}$ applied perpendicular to the air/water interface induces in
each particle a magnetic moment $\vec{M}= \chi \vec{B}$ which leads to
a repulsive dipole-dipole pair-interaction energy of $\beta v(r) =
\Gamma/ (\sqrt{\pi\rho}r)^{3}$ with the dimensionless interaction
strength given by $\Gamma = \beta (\mu_{0}/4 \pi) (\chi B)^{2} (\pi
\rho)^{3/2}$ ($\beta = 1/kT$ the inverse temperature, $\chi$ the
susceptibility). The interaction can be externally controlled by means
of the magnetic field $B$.  $\Gamma$ was determined as in
Ref.~(\cite{zahn99}) and is the only parameter controlling the
phase-behavior of the system. For $\Gamma > 60$ the sample is a
hexagonal crystal \cite{zahn00} (see Fig.~(\ref{fig1})). The sample
was tempered at high interaction strength up to $\Gamma = 250$ deep in
the crystalline phase until a 2D-mono-crystal was observed. We here
analyze three different crystals, from hard to soft ($\Gamma =
250,175,75$), and use for each system about 2000 statistically
independent configurations with approximately 1300 particles, recorded
at equal time intervals ($\Delta t = 2 s$) in a $440 \times 440\: \mu
m$ frame using digital video-microscopy with subsequent
image-processing on the computer.  For each of all $N$ particles in a
given configuration, we determine the displacement $\vec{u}(\vec{R})$
of the particle from its equilibrium position $\vec{R}$.

Using the theory of harmonic crystals \cite{ashcroft}, we now derive
an equation guiding us from the measured displacement vectors
$\vec{u}(\vec{R})$ to the eigenvalues of the dynamical matrix. Let
$D_{\mu,\nu}(\vec{q})$ ($\mu,\nu \in \{ x,y \}$) be the dynamical
matrix \cite{ashcroft}, connected through a Fourier transformation to
the matrix $D_{\mu,\nu}(\vec{R},\vec{R}^{\prime})$ which is
essentially the matrix of the second derivatives of the pair-potential
$v(r) \sim \Gamma/r^{3}$.  It is obvious that $D_{\mu,\nu}(\vec{q})$
depends linearly on the interaction strength parameter $\Gamma$;
therefore we write $D_{\mu,\nu}(\vec{q}) = (kT \Gamma/a^{2})
\tilde{D}_{\mu,\nu}(\vec{q})$ and obtain the dimensionless dynamical
matrix $\tilde{D}_{\mu,\nu}(\vec{q})$ which is independent of $\Gamma$
($a$ is the lattice constant of a hexagonal lattice).  From the
measured $\vec{u}(\vec{R})$ we seek to determine the eigenvalues of
$\tilde{D}_{\mu,\nu}(\vec{q})$ which we denote by
$\lambda_{s}(\vec{q}) a^{2}/(kT \:\Gamma)$. Here the polarization
subscript $s$ stands for the longitudinal ($s=l$) and transversal
($s=t$) mode. The harmonic potential energy of the crystal can be
written in the following form \cite{ashcroft}
\begin{equation}
  \label{eq:1}
  U = \frac{1}{2V} \sum_{\vec{q},\mu,\nu} u_{\mu}^{\ast}(\vec{q})
D_{\mu,\nu}(\vec{q})u_{\nu}(\vec{q})\:
\end{equation}
with $V=N v_{0}=N \sqrt{3}a^{2}/2$ and $u_{\nu}(\vec{q})$ being the $\nu$'th
component of the Fourier transform of the displacement vectors
$\vec{u}(\vec{R})$.  The equipartition theorem for classical harmonic
Hamiltonians states that on average every mode has an energy of
$kT/2$. Thus $(1/2V) \langle
u_{\mu}^{\ast}(\vec{q})D_{\mu,\nu}(\vec{q})u_{\nu}(\vec{q}) \rangle = kT/2$
and this leads us to \cite{chaikin}
\begin{equation}
  \label{eq:2}
  \frac{1}{V}
\langle u_{\mu}^{\ast}(\vec{q}) u_{\nu}(\vec{q}) \rangle
= kT D^{-1}_{\mu,\nu}(\vec{q})\:
\end{equation}
where in our case the average has to be taken over all measured
configurations. Introducing with $p_{s} (\vec{q})$ ($s=t,l$) an
abbreviation for the eigenvalues of the matrix $\Gamma \langle
u_{\mu}^{\ast}(\vec{q}) u_{\nu}(\vec{q}) \rangle /(V a^{2})$, one
arrives at
\begin{equation}
  \label{eq:3}
\frac{1}{p_{s}(\vec{q})} =
\frac{\lambda_{s}(\vec{q}) a^{2}}{kT \:\Gamma}\:,
\end{equation}
($s=t,l$).

Static and slowly moving distortions of the lattice are the main
source of error in our experiment. We sometimes observe long-range
bending of lattice lines, a finite-size problem which in soft crystals
can be partly overcome in giving the crystal enough time to
equilibrate. This takes the more time the harder the crystal is. For
our hardest crystal we have not managed to avoid a small, but clearly
visible bending of lines. A second problem is related to the
determination of each particle's equilibrium position $\vec{R}$,
without which the displacement vectors $\vec{u}(\vec{R})$ cannot be
determined. A cooperative drift of all particles can be observed, a
behavior typical of 2D crystals. An illustrative example pictures can
be found in \cite{zheng} and also in \cite{zahn00}: particles can
depart significantly from their lattice sites, but keep a nearly
constant distance from each other so that nearby trajectories are
similar.  However, when resorting to a local coordinate system
\cite{zheng} the overall symmetry remains crystal-like, and the
crystal melts in accordance with the Lindeman criterion in spite of
the collective drift (measured in the global coordinate system, the
root mean square displacement would diverge at long times, indicating
the instability of the 2D crystal \cite{Bedanov}). These observations
have been confirmed experimentally \cite{zahn00}.  To correct our data
for drift, we first calculated coarse-grained trajectories by
averaging over a sliding time window $\Delta T$ having a width of 25,
40 and 60 $\Delta t$ for $\Gamma = 250,175$ and $75$, respectively.
We then analyzed the short-time displacement of the particles with
respect to these coarse-grained trajectories to obtain the true
fluctuations of the underlying crystal. Our whole data evaluation
procedure has been successfully tested by processing data obtained
from Monte Carlo simulations, using the pair-potential and parameters
of our experiment.

\begin{figure}
\includegraphics[width=0.5 \textwidth]{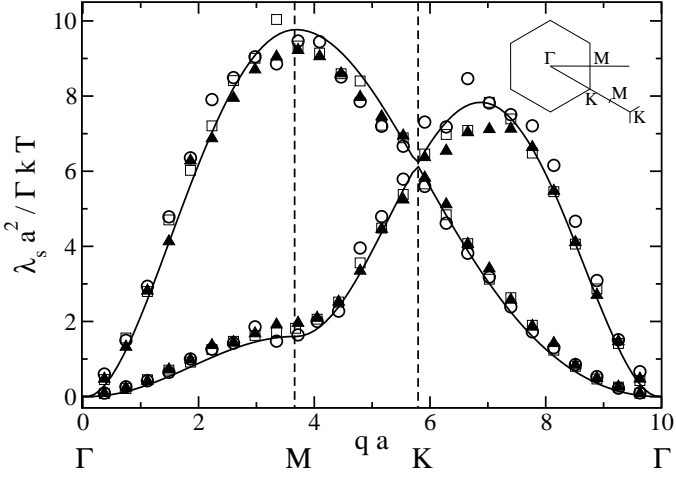}
\caption{\label{fig2}
  Band-structure of harmonic lattice spring constants of a 2D
  colloidal crystal. Symbols for constants experimentally determined
  with eq.~(\ref{eq:2}) and (\ref{eq:3}) from the relative
  displacement of the particles from their equilibrium
  position for a soft ($\Gamma = 75$, empty circles), hard
  ($\Gamma=175$, empty squares) and very hard ($\Gamma=250$, filled
  triangles) crystal; solid lines for the theoretical band-structure
  calculated from standard harmonic crystal theory using the
  pair-potential $\beta v(r) = \Gamma/ (\sqrt{\pi\rho}r)^{3}$.
  The inset shows the first Brillouin zone of the hexagonal lattice
  and labels for high-symmetry points, defining the lines in the
  interior and on the surface of the first Brillouin zone along which
  the band-structure is plotted. The upper curve corresponds to the
  longitudinal, the lower one to the transversal mode.}
\end{figure}

Fig.~(\ref{fig2}) shows $1/p_{s}(\vec{q})$ from eq.~(\ref{eq:3}) as
obtained from the measured set of displacement vectors for $\Gamma =
250$, $175$ and $75$, and compares it to the theoretical
band-structure (solid lines) of a harmonic crystal having a
two-dimensional hexagonal lattice ($a = 12.98$ $\mu m$). The latter is
based on the second derivatives of the known pair-potential and
results from diagonalizing $\tilde{D}_{\mu,\nu}(\vec{q})$
\cite{ashcroft}. 17 neighbor shells have been taken into account in
$D_{\mu,\nu}(\vec{R},\vec{R}^{\prime})$, the difference to the results
for only 3 shells is already tiny. We find good agreement for both the
transversal and longitudinal mode. No fit parameter has been used.
Pre-averaging with a finite time-window improves the agreement.
Without it (i.e. if taking the average over the whole trajectory to
define $\vec{R}$), the peak at the $M$-point in the band-structure is
about 10~\% smaller for each $\Gamma$ than it is in Fig.~(\ref{fig2}).
The data are particularly sensitive to the quality of the crystal near
the edges of the first Brillouin zone, especially near the $M$-point.
The uncertainty in determining the direction of the lattice lines plus
the bending of these lines explain the remaining differences between
the theoretical and the experimental band-structure, but also the
differences between the three different crystals. We also checked for
the occurrence of dislocations in all our samples. Only thermally
activated dislocation pairs have been observed, but no static,
isolated dislocation destroying the crystal symmetry.
\begin{figure}
\includegraphics[width=0.5 \textwidth]{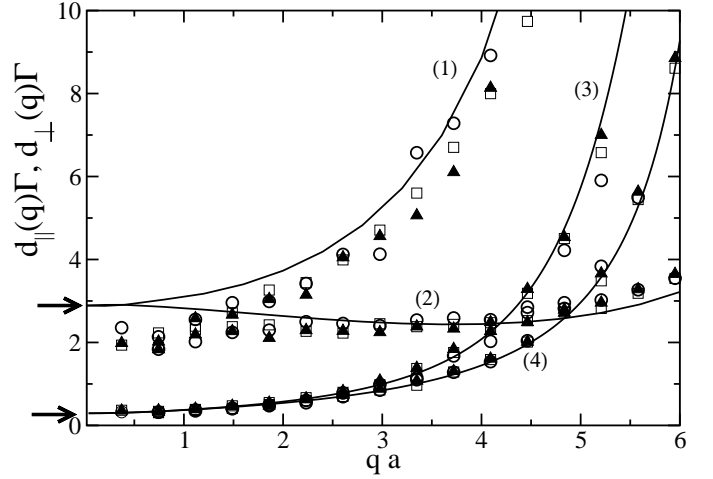}
\caption{\label{fig3}
  $d_{||}(\vec{q}) \Gamma$ and $d_{\bot}(\vec{q}) \Gamma$ from
  eq.~(\ref{eq:4}), evaluated with the same data as taken in
  Fig.~(\ref{fig2}).  For $q \to 0$, $d_{||}(\vec{q}) \to
  kT/((K+\mu)a^{2})$ and $d_{\bot}(\vec{q}) \to kT/(\mu a^{2})$ where
  $K$ is the bulk modulus and $\mu$ the shear modulus. Curves (1) and
  (2) are $d_{\bot}(\vec{q}) \Gamma$ in the $\Gamma \to M$ (curve (1))
  and the $\Gamma \to K$ (curve (2)) direction, (3) and (4) are
  $d_{||}(\vec{q}) / \Gamma$ in the $\Gamma \to K$ (curve (3)) and the
  $\Gamma \to M$ (curve (4)) direction.  Symbols, lines and the labels
  $\Gamma,K,M$ for symmetry-points are all defined in
  Fig.~(\ref{fig2}). Arrows indicate the $T=0$ prediction of the elastic
  moduli, see text.}
\end{figure}

It is instructive to study the $q \to 0$ limit. We define
$\vec{q}_{||}=q_{x}\vec{e}_{x}+q_{y}\vec{e}_{y}$ and
$\vec{q}_{\bot}=-q_{y}\vec{e}_{x}+q_{x}\vec{e}_{y}$ where
$\vec{q}=q_{x}\vec{e}_{x}+q_{y}\vec{e}_{y}$, and multiply on either
side of eq.~(\ref{eq:2}) $\vec{q}_{||}$ and $\vec{q}_{\bot}$ from the
left and the right, to find
\begin{equation}
  \label{eq:4}
d_{i}(\vec{q}) \equiv
\frac{v_{0}}{V} \langle |\vec{q}_{i}\vec{u}(\vec{q})|^{2}\rangle
= kT \sum_{\mu,\nu} q_{i,\mu} D^{-1}_{\mu,\nu}(\vec{q})q_{i,\nu}\:,
\quad i=||,\bot \:.
\end{equation}
It can be shown that $\lim_{q \to 0}d_{||}(\vec{q}) = kT/ ((K+\mu)
a^{2})$ and $\lim_{q \to 0}d_{\bot}(\vec{q}) = kT/(\mu a^{2})$ where
$K$ and $\mu$ are the bulk and shear elastic moduli of continuum
theory.  Since $D_{\mu,\nu}(\vec{q})$ depends linearly on $\Gamma$,
the quantity $\Gamma d_{i}(\vec{q})$ ($i=||,\bot$) is independent of
$\Gamma$.  Fig.~(\ref{fig3}) shows $d_{||}(\vec{q}) \Gamma$ and
$d_{\bot}(\vec{q}) \Gamma$, evaluated with the experimental data for
all three crystals ($\Gamma = 75, 175, 250$, symbols) and
theoretically using the dynamical matrix $D_{\mu,\nu}(\vec{q})$ and
the pair-potential (solid-lines).  For the pair-potential $\sim
\Gamma/r^{3}$ the elastic constants can be calculated to be $Ka^{2}/kT
= 3.461 \Gamma$ and $\mu = K/10$ in the limit $\Gamma \to \infty$
($T=0$) \cite{wille,zahn031}. Arrows in Fig.~(\ref{fig3}) indicate the
prediction of the $T=0$ calculation.  While the data on the
longitudinal branch (curve (3) and (4)) show excellent agreement and
correctly approach the $T=0$ inverse bulk modulus, theoretical and
experimental data on the transversal branch disagree at low $q$.  This
is clearly a finite size effect. Taking data from this very
experiment, it has already been shown \cite{zahn031} that appropriate
finite size scaling leads to an almost perfect agreement with the
$T=0$-prediction of the elastic moduli. We should also remark that the
location of the branching points of the $\Gamma \to K$ and $\Gamma \to
M$ curves in Fig.~(\ref{fig3}) reveal that the assumption of isotropy
is justified only if $qa > 1$ for the transversal and $qa>3$ for the
longitudinal modes. This defines the limits for the continuum approach
often chosen to describe this system.

In summary, we used video-microcopy data to determine the wave-vector
dependent normal mode spring constants of a two-dimensional colloidal
model crystal. We checked the continuum limit and compared the
experimental data to the predictions of the classical theory of a
harmonic crystal.  Our data evaluation procedure can be seen as an
illustration of the validity of the equipartition theorem which we
used to derive the spring constants from the particles' displacement
vectors. The success of our undertaking was not clear from the
beginning; if a crystal in 2D is not stable, how can one measure the
normal mode spring constants?  Here the ideas put forward in
\cite{zheng,zahn00} proved helpful, specifically the introduction of a
local coordinate system.  Analyzing particle distributions at
equilibrium, we were allowed to completely ignore the lattice
dynamics. In this context, it is worth remembering that
Fig.~(\ref{fig2}) is not a phonon-dispersion relation in the classical
sense as there are no phonons propagating with $\omega_{s}(\vec{q}) =
\sqrt{\lambda_{s}(\vec{q})/m}$. Our results suggest that it is more
appropriate to think of a colloidal crystal as a bead-spring lattice
immersed in a viscous fluid \cite{ohshima01}.  A normal vibration mode
then transforms into a 'normal relaxation mode'
\cite{ohshima01,hurd82}, and the motion of a particle is to be
understood as superposition of these 'normal relaxation modes'. A
time-dependent analysis of our data would allow to study the
relaxation process of these normal modes. From the relaxation times
and armed with the spring constants measured here, one could then
proceed to evaluate the $\vec{q}$-dependent friction factors to study,
for example, hydrodynamic forces.  These avenues await further
investigations. We finally remark that the statics and dynamics of
overdamped phonons in two-dimensional colloidal crystals may also be
seen as a contribution to our understanding of surface phonons
\cite{kress}.\\

Stimulating discussions with R. Klein and E.Trizac are gratefully
acknowledged. We also acknowledge financial support from the Deutsche
Forschungsgemeinschaft (European Graduate College 'Soft Condensed
Matter' and Schwerpunktprogramm Ferrofluide, SPP 1104)


\end{document}